\documentstyle[12pt,a4,epsfig]{article}
\newcommand{\gb}{Gauss-Bonnet term}
\newcommand{\be}{\begin{equation}}
\newcommand{\ee}{\end{equation}}
\newcommand{\bc}{\begin{center}}
\newcommand{\ec}{\end{center}}

\newcommand{\hp}{\hspace{.5cm}}

\newcommand{\integrat}{\int d^5x \sqrt{-g}}
\newcommand{\ap}{\alpha'}
\newcommand{\RT}{R_{\mu \nu \rho \sigma}}
\newcommand{\RTU}{R^{\mu \nu \rho \sigma}}
\newcommand{\riccit}{R_{\mu \nu}}
\newcommand{\riccitu}{R^{\mu \nu}}
\newcommand{\g}{g_{\mu \nu}}

\newcommand{\vepsi}{\varepsilon}
\newcommand{\eom}{equation of motion}

\newcommand{\nc}{{1 \over 16 \pi G_5}}
\newcommand{\lb}{\left(}
\newcommand{\rb}{\right)}
\newcommand{\tinte}{\int_0^{\beta}dt}
\newcommand{\V}{V_3}
\newcommand{\vinte}{\int d\Omega_3}
\newcommand{\tinteads}{\int_0^{\beta'}dt}
\newcommand{\rinteads}{\int_{0}^{\tilde R}dr}
\newcommand{\om}{\omega}

\newcommand{\Nc}{16 \pi G_5}
\newcommand{\ob}{\over b^2}

\newcommand{\Rr}{(\tilde R^4 - r_+^4)}

\newcommand{\hhp}{\hspace{.15cm}}
\newcommand{\la}{Lagrangian}
\newcommand{\ds}{\delta \psi}
\newcommand{\sdg}{\sqrt{-g}}

\newcommand{\ls}{{\cal L}_{\xi}\psi}
\newcommand{\hps}{\hspace{.2cm}}

\newcommand{\asch}{AdS-Schwarzschild}
\newcommand{\sectiono}[1]{\section{#1}\setcounter{equation}{0}}
\newcommand{\p}{\partial}
\newcommand{\ben}{\begin{eqnarray}\displaystyle}
\newcommand{\een}{\end{eqnarray}}
\newcommand{\bt}{\bullet}

\begin{document}

{}~
{}~

\hfill\vbox{\hbox{hep-th/0604070}}\break

\vskip .6cm

{\baselineskip20pt
\begin{center}
{\Large \bf
On Euclidean and Noetherian Entropies in AdS Space
}

\end{center} }

\vskip .6cm
\medskip

\vspace*{4.0ex}

\centerline{\large \rm Suvankar Dutta and Rajesh Gopakumar}

\vspace*{4.0ex}

\centerline{\large \it Harish-Chandra Research Institute}

\centerline{\large \it  Chhatnag Road, Jhusi,
Allahabad 211019, INDIA}

\vspace*{1.0ex}

\centerline{E-mail: suvankar@mri.ernet.in, gopakumr@mri.ernet.in}

\vspace*{5.0ex}

\centerline{\bf Abstract}
\bigskip
We examine the Euclidean action
approach, as well as that of Wald, to the entropy of black holes in
asymptotically $AdS$ spaces. From the point of view of holography
these two approaches are somewhat complementary in spirit and it is
not obvious why they should give the same answer in the presence of
arbitrary higher derivative gravity corrections. For the case of the
$AdS_5$ Schwarzschild black hole, we explicitly study the leading
correction to the Bekenstein-Hawking entropy in the presence of a
variety of higher derivative corrections studied in the literature,
including the Type IIB $R^4$ term. We find a non-trivial agreement
between the two approaches
in every case. Finally, we give a general way of understanding
the equivalence of these two approaches.
\vfill \eject

\baselineskip=18pt

\tableofcontents

\sectiono{\large \bf Introduction}\label{s1}

The Bekenstein-Hawking entropy of black holes was one of the first clues
to the holographic nature of gravity. It indicated that any microscopic
accounting of this entropy would entail that the underlying
degrees of freedom are those of a local theory in one lower dimension.
The AdS/CFT conjecture \cite{malda, gkp, witten}
has been a remarkable realisation of this idea,
giving a detailed dictionary between a theory of gravity and
a gauge theory in one lower dimension.

Nevertheless, holography remains quite mysterious as a dynamical
property of a theory with general covariance. To gain a better
understanding of the mechanism giving rise to holography in theories
of gravity, it seems worthwhile to go back to a closer examination
of the entropy of black holes. In this context, the Noetherian (or Wald's)
approach to the calculation of entropy in theories of gravity, with arbitrary
higher derivative corrections to the Einstein-Hilbert action, is
particularly relevant \cite{wald}. For it not only provides a
concrete method to
calculate systematic corrections to the area law entropy, but {\it also}
gives the answer in a form which is holographic in spirit.
The entropy is given in
terms of local quantities evaluated on the so-called bifurcate horizon,
which is a special codimension two surface in the
black hole spacetime.

We might therefore imagine a direct relation of this entropy to the
holographic description afforded in asymptotically anti-de
Sitter spacetimes. In this case, the entropy of a black hole in
AdS space is related to the thermodynamic entropy of the boundary
gauge theory at a finite temperature (which is the same as the
Hawking temperature of the black hole). Thus there are two
apparently different holographic descriptions of the entropy. One in
terms of the horizon and the other in terms of the boundary
gauge theory. In a rough sense, the latter is the UV description from
the microscopic (gauge theory) point of view, while the former is the
IR description, more appropriate from the coarse grained gravity point
of view. One might therefore expect some kind of RG flow to relate the
two.
\begin{figure}
\begin{center}
\leavevmode
\epsfxsize=2.5in
\epsffile{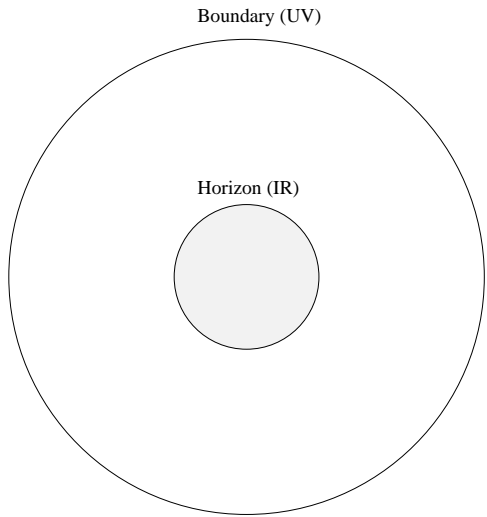}
\caption{RG flow between two holographic descriptions}
\end{center}
\end{figure}
One of the original motivations behind the present work was to try
and make more precise this kind of relation between these two
descriptions. To that end, recall that the AdS/CFT dictionary is
very naturally phrased in terms of a Euclidean functional integral
relation between bulk and boundary quantities\footnote{On the bulk
side the Euclidean functional integral is to be understood only in a
semiclassical sense which we can continue to use in the presence of
higher derivative corrections, i.e. in an expansion about the large
$N$, strong 'tHooft coupling limit.}. In particular, the free energy
(or entropy) of the Euclidean thermal field theory is the same as
the free energy (or entropy) of the gravity configuration as
evaluated by the Euclidean gravitational action. Therefore it is
essential to understand the relation, in the presence of higher
derivative corrections to the Einstein action, between the Euclidean
computation of black hole entropy and Wald's prescription. This
would be a first step in relating the holographic gauge description
to the holographic Wald description.

If both the Euclidean and Noetherian approaches to entropy are to be
sensible prescriptions for obtaining the entropy in a gravity theory
in AdS, then they ought to give the same answer\footnote{In the case of
$AdS_3$, the Euclidean approach was combined
with that of Wald for studying the effect of Chern-Simons terms
 \cite{krlarsen}. See also \cite{sahoosen1, sahoosen2}.}.
However, this is not
immediately apparent. For one, the original derivation by Wald
explicitly made use of the asymptotically flat nature of the space
time. No complete proof seems to have been given yet for the similar
result in asymptotically AdS space \footnote{We have been informed 
by K. Skenderis (private communication) that the
boundary counterterm method employed in \cite{skenderis} can be 
generalised to the higher derivative case and be used to demonstrate
Wald's result for asymptotically locally AdS spaces. The 
boundary counterterm procedure then also provides a link to the 
Euclidean approach.}. 
On the other hand, the Euclidean 
action
prescription in asymptotically AdS space (going back to Hawking and
Page \cite{page}) makes use of a nontrivial background 
subtraction procedure. We
will review this procedure in the next section. For now, we merely
remark that due to the fact that there is a relation between the
sizes of the thermal circles of the background AdS and the black
hole, this is not just a matter of removing an infinite additive
constant, but is crucial to getting the correct finite answer.

To illustrate the non-trivial nature of the agreement between these
two approaches we present some explicit computations of the leading
corrections to the entropy of a five dimensional $AdS$ Schwarzschild
black hole in the presence of a variety of higher derivative
gravitational corrections. The terms we consider are those that
arise as $\alpha^{\prime}$ corrections in string theory. The first
case is the addition of a Gauss-Bonnet term to the Einstein Hilbert
action. This has been studied in detail in the literature (see for 
example \cite{cai} and
references therein) and we use
it as a warmup example . The next case is of a genuinely higher
derivative term $\RT\RTU$. This was studied in \cite{noj, neupane}
and there is some confusion regarding this term.
In \cite{noj},
disagreement was claimed between the Euclidean and Wald's 
expressions even to leading order in $\alpha^{\prime}$. 
\cite{neupane}
obtained a different result 
which actually agrees with Wald's answer (though there is 
an unnecessary remark here that the agreement is only for large
mass black holes).  
We find
complete agreement between the Euclidean and the Wald approaches, to
this order. Finally, we consider the leading $R^4$ correction that
arises in Type IIB supergravity, which plays an important role in
the AdS/CFT context. For this term, the authors of \cite{gubser} had
already studied the correction to the entropy of the planar AdS
Schwarzschild black hole in the Euclidean approach. Here we apply
Wald's prescription to this term and find exactly the same
correction as obtained by \cite{gubser}. In fact, we
also find agreement in the case of the usual AdS Schwarzschild black
hole (i.e. not just in the large mass or planar limit). As we will
see, in all cases, the agreement is not at all manifest from the
prescriptions themselves.

Prompted by this agreement and motivated by our desire to
directly connect the two approaches, we present an argument to
understand this
equivalence. We use the Noetherian definition of
mass in asymptotically AdS spaces given 
in \cite{marolf}\footnote{See 
relatedly, \cite{skenderis}
where things are worked out for the two derivative Lagrangian. 
\cite{koga, koga2}
attempt to generalise some of the considerations of \cite{marolf}
to a class of higher derivative Lagrangians.}
and show that
the subtraction scheme used to define the entropy in the Euclidean approach
arises quite naturally in the Wald approach.
For simplicity, we restrict to static cases.

The organisation of this paper is as follows. In the next two sections
we give a quick overview of the Euclidean and the Wald methods respectively.
Then we present in Section 4, the computations for the individual
cases of higher derivative corrections. In section 5, we
give the argument for why the two methods should give the same
results. The last section carries some of the conclusions.
In a first appendix, we illustrate the Hamiltonian definition of
mass in AdS with a specific example. In the second appendix, we argue that
the the Euclidean definition of mass must agree with the Hamiltonian one.


\sectiono{\bf A Brief Overview of the Euclidean Method}

In this section and the next we will give a quick overview of the
two different approaches to calculating the entropy of black holes in
asymptotically Anti-de Sitter space. We start with the Euclidean
approach 
\footnote{A somewhat different Euclidean
prescription to compute the black hole
entropy is given in
\cite{Solodukhin}} where a precise form of subtraction is
important in getting the correct energy and entropy  
of AdS black holes. We will illustrate this procedure through the case 
of the five dimensional AdS-Schwarzschild black hole.

The Euclidean $AdS_5$ metric is given by,
 \be \label{e2.1}
ds_{AdS}^2=\lb 1+{r^2 \over b^2}\rb d\tau^2+{dr^2 \over \lb
  1+{r^2 \over b^2} \rb}+r^2d\Omega_3^2.
\ee
Here $b$ is the radius of $AdS$ space. In these coordinates the
Euclidean time $\tau$ of the anti-de Sitter space time can be taken
to be periodic with arbitrary period {\bf $\beta'$}. This is then the
Euclidean geometry describing thermal anti-de Sitter space, with
inverse temperature {\bf $\beta'$}.

In these same coordinates, the Euclidean $AdS_5$ Schwarzschild metric is given
by,
\be \label{e2.2}
ds_{BH}^2=\lb 1+{r^2 \over b^2}-{\omega M \over
r^2}\rb d\tau^2+{dr^2 \over \lb
  1+{r^2 \over b^2}-{\omega M \over r^2} \rb}+r^2d\Omega_3^2.
\label{eq:adssh}
\ee
Where,
\be
\omega={\Nc \over 3V_3}
\ee
with $V_3$ being the volume of the unit three sphere.
This space time has a horizon at $r=r_+$, given by,
\be \label{e2.3}
1+{r_+^2 \over b^2}\,-\,{\omega M \over r_+^2}\,=\,0.
\label{eq:rplus}
\ee
The apparent singularity at $r=r_+$ is just
like the singularity at the origin in the polar coordinate system
and can be removed by regarding $\tau$ as an angular variable with
period,
\be \label{e2.4}
\beta={2 \pi b^2 r_+ \over 2r_+^2 +
b^2} \equiv \beta_0. \label{eq:adsbhtemp}
\ee
Therefore, unlike pure AdS space where the time
coordinate can have an arbitrary period, that of the AdS black hole has
a well defined periodicity, for a given mass. This is identified with the
(inverse) Hawking temperature of the black hole.

Both these metrics arise as solutions to the equation of motion
\be \label{e2.5}
R_{\mu \nu}=-{4 \over b^2}\g
\ee
which follows from the Einstein Hilbert action.
\be \label{e2.6}
I_{EH} =  - {1\over 16 \pi G_5}\integrat \left(R+{12 \over
  b^2}\right).
\ee
In what follows we will consider higher derivative corrections
to the Einstein-Hilbert Lagrangian. These will lead, in particular,
to systematic
corrections to the black hole metric, Eq. \ref{eq:adssh}, which we can compute
in a perturbative expansion in the coefficients of the additional terms.
We are interested in the corrections to the thermodynamics of the black hole
spacetime from these terms.

Therefore, let us first
discuss the thermodynamics of black holes in the Euclidean
framework as generally prescribed by Gibbons and Hawking \cite{gibbons}.
The canonical
partition function is defined by a
functional integral over metrics
with the Euclidean time coordinate $\tau$ identified with period $\beta$,
defined above.
\be
Z=\int [{\cal D}g] e^{-I_E}
\ee
$I_E$ is the Euclidean action which could have the Einstein-Hilbert piece
as well as higher derivative corrections.
In the semi-classical limit that we are
considering, the dominant contribution to the path
integral comes from classical solutions to the
equations of motion. In this case,
\be
\ln Z = -I_E^{cl}.
\ee
Thus the action $I_E^{cl}$,  evaluated on solutions, is proportional to the
free energy.
Therefore the energy (or mass) of the black hole is given by,
\be \label{eq:euclieng}
E=-{\p \over \p \beta}\ln Z = {\p I_E^{cl} \over \p \beta}
\label{eq:euclmass}
\ee
and the entropy of the black hole is given by
\be \label{eq:euclient}
S=\beta \, E - I_E^{cl}.
\label{eq:euclentr}
\ee

For asymptotically AdS black holes, the prescription needs to be
modified. This is suggested by the fact that if we
calculate the action $I_E^{cl}$ on a solution to the equation
of motion, it turns out to be infinite. So we need to specify a
subtraction procedure which will give sensible answers.
To get the right results one subtracts the contribution of global AdS
after a suitable regularisation.
However, one has to be careful in doing this \cite{page}.
There are two points to be kept in mind:
{\bf (i)} For the black hole spacetime we evaluate the action integral
only in the region $r_+ \leq r \leq \tilde{R}$
where $\tilde{R}$ is an IR cutoff on the spacetime
and $r_+$ is the location of the horizon. Whereas in the
AdS space time the region of integration
is $0 \leq r \leq \tilde{R}$.
{\bf (ii)} The crucial point, however, is that we do not
take the black hole spacetime and the reference AdS
spacetime as having the same periodicity in the time direction. The black hole
spacetime has a fixed periodicity $\beta$ (for instance given by (2.8)
in the case of the AdS-Schwarzschild metric).
One adjusts the period $\beta'$ of the globally AdS spacetime
such that the geometry at the
hypersurface $r=\tilde{R} $ is the same in both cases \cite{page}, i.e,
\be \label{e2.9}
\beta [g_{\tau\tau}^{BH}(r=\tilde{R})]^{1/2}
=\beta' [g_{\tau\tau}^{AdS}(r=\tilde{R})]^{1/2}.
\label{eq:temprel}
\ee
Through this relation,
the periodicity of the reference AdS spacetime depends on the
parameters (such as the mass etc.) of the black hole spacetime.

Let us illustrate how this regularisation, specified above, works in
the simplest case of the Einstein-Hilbert action \cite{page}.
This will also show how the subtraction does not just
remove the divergent piece but is also important in getting the finite
contributions to the mass and entropy correct.

Evaluating the Einstein-Hilbert action on the solution Eq.\ref{eq:adssh},
we get,
\be
I_{BH}= {2\beta \over 3\omega b^2}(\tilde R^4 -r_+^4).
\ee
As prescribed, the radial integration has been carried out
from $r_+$ to $\tilde R$, with
$\tilde R \rightarrow \infty$. We can then
calculate the mass using Eq. \ref{eq:euclmass} together with the
relations Eqs.\ref{eq:rplus} and \ref{eq:adsbhtemp}. We obtain
\footnote{Here we have chosen to write the expression in
a particular combination which will be used later. The main
point to notice is that the finite part of $\p I / \p \beta$ does {\it not}
give the mass of the black hole.}
\be
{\p \over \p \beta}I_{BH}=
{2\over 3\omega b^2}\tilde R^4 + {2M \over 3} -{2\over 3\omega b^2}
{r_+^2(2r_+^2+b^2)^2 \over (b^2-2r_+^2)}.
\label{eq:Esch}
\ee
The first term is the divergent piece while the rest are finite.
We can also naively calculate the entropy using Eq. \ref{eq:euclient}.
\ben
\tilde{S}&=& \beta {\p \over \p \beta} I_{BH} - I_{BH} \nonumber \\
&& =\lb {V_3 r_+^3 \over 4G_5}\rb {4r_+^2 \over 2r_+^2
-b^2}. \label{eq:Ssch} \een
Note that the entropy as calculated with
the above cutoff, gives a {\it finite} answer. The divergent pieces
have canceled out. But this finite piece is {\it not} the correct
area entropy which one expects for the Einstein-Hilbert action. As
we will see, we obtain the right answer only after the prescribed
subtraction of the reference AdS geometry.

Similarly one can evaluate the action on the (regularised) AdS spacetime
\be
I_{AdS}={2\beta'\over 3\omega b^2}\tilde R^4.
\ee
Here the radial
integration has been done from zero to $\tilde R$. $\beta$ and $\beta'$
are related by Eq. \ref{eq:temprel}. For large $\tilde R$ we get,
\be
\beta'=\beta \lb 1 -{\omega M b^2 \over 2 \tilde R^4}, \rb
\ee
where $\omega M$ is a function of $r_+$ given by Eq. \ref{eq:rplus}.
We then find
\be
{\p \over \p \beta}I_{AdS} = {2\over 3\omega b^2}\tilde R^4 - {M \over 3} -
{2\over 3\omega b^2}{r_+^2(2r_+^2+b^2)^2 \over (b^2-2r_+^2)}.
\label{eq:Eads}
\ee
Subtracting Eq.\ref{eq:Eads} from Eq.\ref{eq:Esch}
we find that the energy of the black hole is exactly M!
Similarly, for the entropy, we evaluate
\be
\beta{\p \over \p \beta}I_{AdS} - I_{AdS}
= -{4\pi b^2\over 3\omega b^2}{r_+^3 (2r_+^2+b^2) \over b^2 -2r_+^2}.
\label{eq:Sads}
\ee
We need to subtract Eq. \ref{eq:Sads} from Eq. \ref{eq:Ssch} to get the
entropy
\be
S=\lb {\V r_+^3 \over 4\, G_5}\rb = {A_{BH} \over 4 G_5}.
\ee
Thus we get the expected answer from the nontrivial interplay between
the actions evaluated on the AdS Schwarzschild geometry and the background
AdS spacetime.


\sectiono{A Brief Overview of Wald's Approach}

The essence of Wald's approach consists of three steps. The first is
to give a general expression for Noether currents and charges
corresponding to arbitrary diffeomorphisms. The next step is to
use this to construct Hamiltonians corresponding to these Noether charges.
In special cases, these Hamiltonians are related to the usual
conserved charges such as mass, angular momentum etc. and are given by
surface integrals at infinity on a Cauchy surface.
Finally, in the case of
certain killing vectors such as for time translation
invariance one can relate the variation of the Hamiltonian
(corresponding to say, energy)
to that of the  Noether charge evaluated on the
bifurcate horizon. This relation can then be interpreted as the
first law of black hole thermodynamics with
the latter quantity identified as
being proportional to the entropy. We will now give a
quick review of these steps, referring the reader to the
original papers \cite{wald, iyer} for more details.

\subsection{Construction of Noether charges}

Let {\bf L} be a diffeomorphism invariant Lagrangian, in $n$ spacetime
dimensions built out of
the metric and other fields, collectively denoted by $\psi$
\footnote{We will follow
Wald's convention of viewing the Lagrangian as a top form rather
than a scalar density.
Similarly, other tensor densities will also be dualised and viewed
as appropriate
forms. To distinguish these from the usual densities, we will denote
the forms in boldface.}.
Under
any arbitrary field variation $\delta \psi$ the \la\ varies as,
\be
\delta {\bf L}= {\bf E}(\psi)\delta\psi
+d{\bf \Theta}(\ds).
\ee

The classical equations of motion are given by ${\bf E}=0$. The second term
is a boundary term which can depend on both $\ds$ and its derivatives.
The notation is abbreviated here so that a sum over the tensor indices,
for instance for $\ds$, is understood.

For instance, for
a generally covariant Lagrangian which is a function of the metric and
different powers of the Riemann tensor, but {\it not} covariant derivatives
of the Riemann tensor, ${\bf \Theta}$ can be chosen to be of the form
\be
\Theta^a(\delta g)=2 E_R^{abcd} \nabla_d \delta g_{bc} -
2\nabla_d E_R^{abcd} \delta g_{bc},
\ee
where
\be
E_R^{abcd}={\partial L \over \partial R_{abcd}}.
\ee

We will actually be interested in  diffeomorphisms,  where the field
variations are given by the Lie derivative $\ds={\cal L}_{\xi} \psi$
($\xi^a$ is the infinitesimal generator of a diffeomorphism). The
resulting variation of a covariant {\la} is then a total derivative.
\be
\delta {\bf L}  = {\cal L}_{\xi} {\bf L}= d(\xi\cdot {\bf L}).
\ee
Here the `$\cdot$' denotes the usual contraction of a vector field with a
form.

Thus, in this case we can define a current
\be
J_{\xi}^a=\Theta^a (\ls) - \xi^a L.
\label{eq:noeth}
\ee
Or equivalently in terms of $(n-1)$ forms,
in the dualised notation
\be
{\bf J}_{\xi}={\bf\Theta}(\ls) - \xi\cdot{\bf L}.
\label{eq:noeth2}
\ee
It satisfies
\be
d{\bf J}_{\xi} = -{\bf E}\ls .
\ee
So $J_{\xi}^a$ is a Noether current which is
conserved for {\it any} $\xi$ when the
equations of motion are satisfied.

Now, for any conserved $J_{\xi}$
there (locally) exists a
$(n-2)$ form ``Noether Charge''
${\bf Q}_{\xi}(\psi)$ constructed out of fields
$\psi$ and $\xi^a$ \cite{lee},
such that whenever $\psi$ satisfies the {\eom},
we have,
\be
{\bf J}_{\xi}=d{{\bf Q}_{\xi}(\psi)}.
\label{eq:jqrel}
\ee
In what follows we will always take $\psi$ to be a solution
of the equations of motion.

\subsection{Hamiltonians and Noether Charges}

As mentioned above, the second step is to relate these Noether charges
to Hamiltonians generating the diffeomorphisms $\xi$ 
(see also \cite{zoupas}).
The essential point here is that the
boundary term ${\bf\Theta}(\ds)$ acts as a ``symplectic potential'' which
enables one to define a symplectic form on the phase space of field
configurations. Considering two arbitrary variations $\delta_1\psi$
and $\delta_2\psi$, we can define a symplectic current ${\bf \omega}$ by
\be
{\bf \omega}(\delta_1\psi, \delta_2\psi)=\delta_2{\bf\Theta}(\delta_1\psi)
-\delta_1{\bf\Theta}(\delta_2\psi).
\ee
The symplectic form (on the space of variations $\delta\psi$)
itself is then defined by
an integral over a Cauchy surface $\Sigma$
\be
\Omega(\delta_1\psi,
\delta_2\psi)=\int_{\Sigma} {\bf\omega}(\delta_1\psi, \delta_2\psi).
\ee

So now an arbitrary variation $\delta\psi$ of the Noether current
can be re-expressed in terms of the symplectic form
\ben
\delta {\bf J}_{\xi} &=&\delta{\bf\Theta}(\ls)-\xi\cdot\delta{\bf L}
\nonumber \\
&=& {\bf\omega}(\delta\psi, \ls)+d(\xi\cdot{\bf \Theta}(\delta\psi)).
\label{eq:omegaid}
\een
In the second equality, we have used the fact that $\psi$ is a solution
to the equations of motion as well as an identity for Lie derivatives.
Integrating Eq.\ref{eq:omegaid}
over the Cauchy surface and since the Hamiltonian for a
vector field $\xi$ is defined via
\be
\delta H_{\xi}=\Omega(\delta\psi, \ls)
\ee
we have
\ben
\label{eq:hamq}
\delta H_{\xi} &=&
\int_{\Sigma}\delta{\bf J_{\psi}}-\int_{\partial\Sigma}
\xi\cdot{\bf \Theta}(\delta\psi) \nonumber\\
&=& \int_{\partial\Sigma}(\delta {\bf Q}_{\xi} -\xi\cdot{\bf
\Theta}(\delta\psi)).
\een
In the second equality we have evaluated
the Hamiltonian on-shell, where 
${\bf J_{\psi}}=d{\bf Q_{\psi}}$. Furthermore, if
the boundary term arises from the variation of a boundary term in
the action,
\be
{\bf\Theta}(\delta\psi)=\delta {\bf B},
\ee
then the above
equation can be integrated to obtain a Hamiltonian $H_{\xi}$. If
$\Sigma$ has only an asymptotic boundary at infinity then the above
Hamiltonian is expressed purely as a surface integral at infinity.
In the asymptotically flat case the Hamiltonian corresponding to an
asymptotic time translation vector field was shown to be equal to
the ADM mass in the case of the Einstein-Hilbert Lagrangian. We will
discuss the asymptotically AdS case later.

\subsection{Entropy as Noether charge}

We can now apply these considerations to {\it killing} vector fields. In
particular, we will specialise to killing vector fields $\xi_H^a$
which are {\it null} on the
codimension one horizon and {\it vanish} on the codimension two
surface called the
bifurcate horizon ${\cal H}$ of a black hole spacetime.

To obtain the first law of black hole thermodynamics, we first
consider a variation $\delta\psi$ in Eq.\ref{eq:omegaid}, which
satisfies the linearised
equations of motion. Since we are
considering a Killing vector field, for which $\ls=0$, we also have
${\bf\omega}(\delta\psi,\ls)=0$. We
use this and integrate both sides of Eq. \ref{eq:omegaid} over a
spatial hypersurface $C$ of the black hole spacetime which has
${\cal H}$ as its interior boundary (in addition to asymptotic
infinity). Since $\ds$ satisfies the equations of motion,
we have $\delta {\bf J} =d(\delta {\bf Q})$. The integrand on both sides
reduce
to total derivatives. The integral thus gets contributions from only
the interior
boundary and from infinity.

For instance, we might take
\be
\xi_H^a = {\partial \over
\partial t}
\ee
in a static black hole background spacetime. In
such a case, using Eq.\ref{eq:hamq}, the outer boundary
contribution gives exactly the change
in the energy or mass.
Thus we get a relation of the form
\be
\delta \int_{{\cal H}}
dS_{ab} \sdg  Q^{ab}= \delta {\cal E}. \label{eq:dqde}
\ee
It turns out to be consistent to make an identification
with the entropy $S$ via
\be
\label{noethent}
{\kappa \over 2\pi}\delta S = \delta \int_{{\cal
H}} dS_{ab} \sdg Q^{ab},
\ee
where $\kappa$ is the surface
gravity (which is constant over ${\cal H}$) and is proportional
to the temperature.
Therefore Eq.\ref{eq:dqde} is the first law of black hole thermodynamics,
\be
\label{eq:istlaw}
{1\over \beta}\delta S = \delta {\cal E}.
\ee

Moreover, from the identification with the Noether charge in Eq.
\ref{noethent}, Iyer and  Wald \cite{iyer} also found a simple expression
for the entropy.
Using the definition of the Noether charge and with the Noether current
given by Eq.\ref{eq:noeth}, they
expressed the entropy in terms of the Lagrangian as
\be
S=-2\pi \int_{{\cal H}} {\delta L
\over \delta R_{abcd}}\epsilon_{ab}\epsilon_{cd}.
\label{eq:wald}
\ee
where $\epsilon_{ab}$ is the binormal to the surface ${\cal H}$.
In the special case where the Lagrangian does not depend on
derivatives of the Riemann tensor, we have
\be
S=-2\pi \int_{{\cal
H}} {\partial L \over \partial R_{abcd}}\epsilon_{ab}\epsilon_{cd}.
\label{eq:wald2}
\ee
It is this expression that we will have occasion to use in the next section in
comparing with the Euclidean answers.

\sectiono{\bf Correction to the entropy from higher derivative
  terms}

In this section we will consider some explicit examples of higher derivative
corrections to the Einstein-Hilbert action
(with negative cosmological constant). The examples are all ones
that arise as $\alpha^{\prime}$ corrections in various string theory effective
actions. They include {\bf a)} the Gauss-Bonnet term,
{\bf b)} $\RT\RTU$ term and {\bf c)} the Type IIB $R^4$ or equivalently
$(Weyl)^4$ term. In all these cases we calculate the leading correction
to the entropy of the black hole geometry which is asymptotically AdS.
Keeping in mind the application to holography and the AdS/CFT conjecture,
we restrict ourselves to the case of five dimensional spacetime.

The calculations are done using both the Euclidean approach as well as
using Wald's formula. In the Euclidean approach this requires one to
correct the leading AdS-Schwarzschild metric and then evaluate the
action and thus entropy according to the prescription outlined in Sec.2.
Similarly, in using Wald's formula, we need to evaluate the
corrections to the area law entropy coming from the additional
contributions to Eq.\ref{eq:wald}. We find agreement between the two separate
calculations in every case, unlike the claims in the literature
to the contrary. As mentioned in the Introduction, the agreement between
these very different seeming modes of calculation is quite surprising.
These explicit checks therefore gives one mutually reinforcing confidence
in both approaches.


\subsection{The Gauss-Bonnet term} \label{sec3.1}

The Gauss-Bonnet term is a very natural correction term to the Einstein
action. It has the feature that despite being built out of terms which
individually give rise to higher than two derivative equations of motion,
the full equation of motion actually has only two derivatives.
This term arises in both the heterotic and bosonic string theory
low energy effective actions, after a suitable field redefinition.

In this case exact black hole solutions are known (see for e.g.
\cite{cai}). However, in keeping with the other cases where we do not have
this luxury, we will work only to leading order in this term.
The action containing the \gb\ is
\be \label{eq:gbacn}
I = - {1\over 16 \pi G_5}\integrat \left(R+{12 \over
  b^2}\right)-{\ap/4 \over  16 \pi G_5}\integrat L_{G.B.}
\ee
where,
\be \label{e3.1.2}
L_{G.B.}=\RT \RTU - 4 \riccit \riccitu + R^2.
\ee
The particular coefficient $\alpha' /4$ is chosen so that $I$ matches with the
low energy effective action of heterotic string theory.

The equation of motion for this action is,
\be
\riccit - {1 \over2}R \g -{6 \over b^2}\g = {\ap \over 8}g_{\mu \nu}
L_{G.B.} -{\ap \over 2}
\left(R_{\mu \rho \sigma \delta} R_{\nu}^{\rho \sigma \delta}
- 2 R^{\rho \sigma} R_{\mu \rho \nu \sigma} -2 R_{\mu}^{\rho} R_{\nu \rho}
+ R \riccit \right),
\label{eq:gbeom}
\ee
from which we get,
\be
R = -{20 \over b^2} -{\ap \over 12}L_{G.B}.
\label{eq:gbricci}
\ee
\noindent
$\bt$ {\bf Correction to the Metric}

We are
interested in treating the Gauss-Bonnet term as a perturbation and
finding the leading correction in $\ap$ to the AdS Schwarzschild
metric. We will therefore look for a solution of the spherically
symmetric, static form
\be
ds^2=B(r)d\tau^2+ A(r)dr^2 +
r^2d\Omega_3^2 \label{eq:sphmet}
\ee
where,
\be
A(r) =A_0(r)(1+ \ap \mu(r)), \hspace{.5cm}
  \hspace{.5cm}
A_0(r)= \left(1+{r^2 \over b^2} - {\omega M \over r^2}\right)^{-1}
\label{apert}
\ee
and
\be
B(r)=B_0(r)(1+\ap \vepsi(r)),\,\,\, \,\, \, \, \, \,\,
  B_0(r) = \left(1+{r^2 \over b^2} - {\omega M \over r^2}\right).
\label{bpert}
\ee
We can solve the equations of motion Eq.\ref{eq:gbeom} keeping in mind that we
can use the unperturbed metric in evaluating the terms proportional to $\ap$.
We obtain
\be
\mu(r)=-A_0(r)
\left({r^2 \over 2b^4} + {M^2 \omega^2 \over 2r^6} \right); \hspace{1cm}
\vepsi(r) = -\mu(r) .
\ee
So that
\be
A(r)\, =\, B(r)^{-1}\, =\, \left(1+{r^2 \over
b^2}-{\omega M \over r^2}+ \ap{r^2 \over 2b^4}+ \ap{\omega^2 M^2
\over 2r^6}\right)^{-1}.
\label{eq:gbmet}
\ee
This solution matches
with the solution found in \cite{cai}\footnote{The parameter $M$
which appears in this solution is {\it not} the mass in the presence of
this correction. In comparing with \cite{cai}, this has to be kept
in mind.}.
\newpage
\noindent
$\bullet$ {\bf Correction to the Black Hole Temperature}

The Euclidean time direction has a periodicity $\beta$ fixed by requiring the
geometry to be smooth at the horizon.
For a spherically symmetric metric of the form in Eq. \ref{eq:sphmet},
the inverse temperature of the black hole is therefore
\be
\label{eq:bhtemp}
\beta={4 \pi \over B'(r_+)}.
\ee
where, $r_+$ is now the (corrected) location of the horizon of the black hole,
$g_{\tau \tau}(r_+)=0$.
\be
\label{eq:gbmasshorizon}
\omega M = r_+^2 \left(1+ {r_+^2 \over b^2}
+ \ap {r_+^2 \over b^4}+{\ap \over 2r_+^2}+{\ap \over b^2}\right).
\ee
Using Eq.\ref{eq:gbmet} we get the corrected inverse temperature,
\be
\label{eq:gbtemp}
\beta={2 \pi b^2 r_+ \over 2r_+^2 + b^2} \left(1+{\ap \over
r_+^2} \right) =\beta_0\left(1+{\ap \over
r_+^2} \right).
\ee

\noindent
$\bt$ {\bf Calculation of Entropy}

Let us write the black hole action \ref{eq:gbacn} to be,
\be \label{eq:gbacnsep}
I_{BH}=-\nc \lb I_0 + I_1 \rb
\ee
where,
\be
I_0= \integrat \left(R+{12 \over b^2}\right),\hp and \hp I_1= {\ap
  \over 4}\integrat  \, L_{G.B.}.
\ee

In $I_0$ we substitute the value of $R$ for the perturbed solution.
For this we use Eq.\ref{eq:gbricci}. In
$I_1$ we can use the unperturbed metric, to this order in $\ap$.
So that
\be \label{eq:gb}
L_{G.B} ={120 \over b^4} + {72 \omega^2 M^2 \over r^8}.
\ee
We get,
\begin{eqnarray}
I_0&=&\integrat \lb {-20 \over b^2} +{12 \over b^2} -\ap{10
\over
  b^4} - \ap {6 M^2 \omega^2 \over r^8}\rb \nonumber \\
&&
= -\tinte \rinteads \vinte r^3\left( {8\over b^2} +\ap {10 \over b^4}
 + {6 \ap M^2\omega^2
  \over r^8}\right)\nonumber \\
  &&
= -{\beta \V (\tilde R^4-r_+^4)}\Bigg({2\over b^2} +\ap {5
  \over 2b^4} +  {3\ap M^2\omega^2\over 2\tilde R^4 \ r_+^4}\Bigg),
\end{eqnarray}
where $\V$ is volume of unit 3 sphere, $\V=2\pi^2$.
$I_1$ can be evaluated
using the unperturbed metric. So that
\be I_1= \ap {\beta \V (\tilde R^4 - r_+^4)}\left({15\over 2b^4} + {9M^2
\omega^2 \over 2\tilde R^4 \ r_+^4}\right). \ee \noindent
Hence $I_{BH}$ is given by,
\ben
I_{BH} = {\beta \V \lb \tilde R^4 - r_+^4\rb \over 16 \pi G_5} \lb {2
  \over b^2} - \ap {5 \over b^4} - \ap {3 M^2 \omega^2 \over \tilde
  R^4 r_+^4} \rb.
\een

The next step is to calculate the action of the background $AdS$
spacetime,
\be I_{AdS}= -\nc \lb J_0 + J_1 \rb \ee
where,
$$
J_0=\tinteads \rinteads r^3 \vinte \lb R + {12 \over b^2}\rb \hp and
\hp J_1={\ap \over 4} \tinteads \rinteads r^3 \vinte L_{GB}.
$$
Here $\beta'$ is the periodicity of the time direction of the AdS
space time. These expressions are easily evaluated,
\be
J_0=-\beta' \V \tilde R^4 \lb {2\over b^2} + {5\ap \over 2b^4} \rb
\ee
and
\be
J_1=\ap {15 \beta' \V \tilde R^4 \over 2b^4}.
\ee
To evaluate $J_0$, we have used the perturbed $AdS$ metric which is obtained
from Eq.\ref{eq:gbmet} by setting $M=0$. As before $J_1$ is evaluated using
the leading order solution.
So $I_{AdS}$ is given by,
\be
I_{AdS}={\beta' \V \tilde R^4 \over \Nc} \lb {2 \over b^2}-  \ap
{5 \over b^4}   \rb.
\ee
The difference between the
AdS-Schwarzschild action and the AdS action is
\begin{eqnarray}
\displaystyle
\label{eq:gbsubtractacn1}
\Delta I &=& I_{BH} - I_{AdS} \nonumber \\
&=&
 -{\V \over 16\pi G_5} \Bigg[ {2\over b^2}
\tilde R^4 (\beta' - \beta) +  {2\over
    b^2} \beta r_+^4 \nonumber \\
&&
 +\ap \lb {3 \beta M^2 \om^2 \over r_+^4} - {5
    \beta r_+^4 \over b^4} - {5 \over b^4}\tilde R^4 (\beta' -\beta) \rb
  \Bigg].
\end{eqnarray}
Since at the outer boundary hypersurface $r={\tilde R}$,
the geometry of the $AdS$-Schwarzschild spacetime
and $AdS$ is the same,
$\beta$ and $\beta'$ are related by Eq.\ref{eq:temprel},
\be
\beta' \lb 1+{\tilde R^2 \over b^2} +
\ap {\tilde R^2 \over 2b^4}\rb^{1/2} = \beta
\lb 1+{\tilde R^2 \over b^2} -
{\om M \over \tilde R^2}+ \ap {\tilde R^2 \over 2b^4} + \ap
    {\om^2 M^2 \over 2\tilde R^6}\rb^{1/2}.
\ee
In other words,
\be
 \beta' = \beta\lb 1 -{1 \over 2 \tilde R^4} \om M b^2
+ \ap {\om M \over 4 \tilde R^4} \rb.
\ee
Substituting $\beta'$ in terms of $\beta$ in equation
\ref{eq:gbsubtractacn1} we get,
\be \label{eq:gbsubtractacn}
\Delta I=  -{\V \over 16 \pi G_5} \beta \left[ \lb {r_+^4 \over b^2}-r_+^2\rb
  +\ap \lb {8r_+^2 \over b^2} + {5 \over 2}\rb\right].
\ee
Using the relation Eq.\ref{eq:gbsubtractacn}
one can calculate the energy of the black
hole, and it comes out to be,
\ben \label{eq:gbenergy}
E&=& {\partial \Delta I \over \partial \beta} \nonumber \\
&=& {3 \V r_+^2 \over 16 \pi G_5}\lb 1+{r_+^2 \over b^2}
+ {\ap \over 2r_+^2}\rb \nonumber \\
&=& M\lb 1-{\ap\over b^2}\rb.
\een
This relation between the energy and the parameter $M$
agrees in comparing with those of \cite{cai}.
The entropy of the black hole is given by
\be
S=\beta {\partial \Delta I \over \partial \beta} - \Delta I.
\ee
Using equations \ref{eq:gbsubtractacn},
\ref{eq:gbtemp} and \ref{eq:gbenergy} the final corrected
Euclidean entropy of
the black hole works out to be,
\be
S={\V r_+^3 \over 4G_5} \lb 1 + \ap {3 \over r_+^2}\rb.
\ee
\\
One can now compare with the entropy
using Wald's formula. It is given for instance by \cite{myers},
\be
S_{Wald}\, =\,{4 \pi \over \Nc}  \, \int_{\cal H} d^3x \sqrt{h} \left
[ 1 \, + \, {\ap \over 2} \, {\cal R}(h) \right ]
\ee
where {\it h} is the determinant of induced metric on the spherical
horizon and ${\cal
R}(h)=h^{ij}h^{kl}R_{ikjl}$.  ${\cal R}$ can be evaluated using
the unperturbed metric on the sphere
$$
{\cal R} = {6 \over r_+^2}.
$$
So $S_{Wald}$ exactly matches with the Euclidean entropy. Note that
$b^2$ does not appear in the entropy.


\noindent
\subsection{The $R^2$ Term} \label{sec3.2}

The second example we consider involves a correction proportional to
$R_{\mu \nu \rho \sigma} R^{\mu \nu \rho \sigma}$ (which we will
call the $R^2$ term for convenience). This is the first
term that one can add to the Einstein-Hilbert action
which genuinely has higher derivatives
(in contrast to the Gauss-Bonnet term) while also changing the
$AdS$ Schwarzschild solution nontrivially to leading order.
This example has been also studied in \cite{noj}\cite{neupane}.
In \cite{noj}, a discrepancy was claimed between the Euclidean and
Wald expressions for the entropy, even to leading order. 
Our results here agree with \cite{neupane} 
who also found that the two methods actually yield the same
result for the leading correction in
$\ap$ (but for arbitrary mass black holes).

We will take the action containing the $R^2$ correction to be
\be \label{eq:r2acn}
I=-\nc \integrat \left[\lb R + {12 \over b^2} \rb + {\ap \over 4}
  R_{\mu \nu \rho \sigma} R^{\mu \nu \rho \sigma}\right].
\ee
Here we have again chosen the coefficient of the higher derivative
term to be $\ap/4$ as in section \ref{sec3.1}.
The equation of motion is given by,
\begin{eqnarray}\displaystyle
\riccit -{R \over 2}\g-{6 \over b^2}\g &=&-{\ap \over 2}
\bigg[ R_{\mu}^{\rho \sigma \delta} R_{\nu \rho \sigma
    \delta}+2\Box \riccit - {1 \over 2} (\nabla_{\mu} \nabla_{\nu}
  +\nabla_{\nu}\nabla_{\mu})R \nonumber \\
&&
 +2R^{\rho \sigma} R_{\mu \rho \nu \sigma}
  - 2 R_{\mu}^{\rho} R_{\nu \rho}\bigg] + {\ap \over 8}\g R_{\alpha \gamma
  \rho \sigma}R^{\alpha \gamma
  \rho \sigma}.
\label{r2eom}
\end{eqnarray}
For the $AdS$ Schwarzschild metric, this leads to a correction at
order $\ap$. In the right hand side of Eq.\ref{r2eom},
we substitute the unperturbed metric to obtain
\be
\riccit + {4 \over b^2} \g = -{\ap \over 2}\left[J_{\mu \nu} -\lb {20
    \over 3b^4}+{12 M^2 \om^2 \over r^8}\rb \g\right].
\ee
Where $J_{\mu \nu}$ is given by,
$$
J_{\mu \nu}=R_{\mu}^{\rho \sigma \delta}R_{\nu \rho \sigma \delta}.
$$
evaluated on the metric \ref{eq:adssh}.
We also record for later,
\be
R=-{20 \over b^2} - \ap \lb{10 \over 3b^2} + {6 M^2 \om^2 \over
r^8}\rb.
\ee
$\bullet$ {\bf Correction to the equation of motion}

We solve the equation of motion
with the spherically symmetric ansatz, Eq. \ref{eq:sphmet},
\be
ds^2=B(r)d\tau^2+ A(r)dr^2 + r^2d\Omega_3^2.
\ee
Using the parametrisation Eqs.\ref{apert},\ref{bpert} and solving
for the functions $\vepsi(r)$ and $\mu(r)$ we find
\be
\vepsi(r)={r^2 \over 6b^4B_0}+{M^2 \om^2 \over 2r^6B_0}, \hp
\hp \mu(r)=-\vepsi(r).
\ee
Hence,
\be
B(r)=A(r)^{-1}=
1+{r^2 \over b^2}-{\om M \over r^2} + \ap \lb {r^2 \over 6b^4} +
  {M^2 \om^2 \over 2r^6} \rb.
\ee
$\bt$ {\bf Correction to the Black Hole Temperature}\\
Using Eq. \ref{eq:bhtemp}
with $r_+$ given by the equation,
\be
\om M =r_+^2+{r_+^4 \over b^2}+{\ap \over 2}\lb 1+ {2r_+^2 \over b^2}+
{4r_+^4 \over 3b^4}\rb
\ee
the inverse temperature $\beta$ comes out to be,
\be \label{eq:r2temp}
\beta={2 \pi b^2 r_+ \over 2r_+^2 + b^2}
\left[1+{\ap \over 3b^2}{2r_+^4 + 3b^4 + 6b^2
r_+^2 \over r_+^2(2r_+^2 + b^2)}\right]=
\beta_0\left[1+{\ap \over 3b^2}{2r_+^4 + 3b^4 + 6b^2
r_+^2 \over r_+^2(2r_+^2 + b^2)}\right].
\ee
$\bt$ {\bf Calculation of Entropy}\\
The action \ref{eq:r2acn} can be evaluated to be
\be
I_{BH} =-\nc \lb I_0+I_1\rb.
\ee
Here $I_0$ is evaluated using the perturbed solution
\ben
I_0&=& \integrat \lb R + {12 \over b^2} \rb \nonumber \\
&=&-{2 \V \beta \over b^2}(\tilde R^4-r_+^4) -\ap{5 \V \beta \over
  6b^4}(\tilde R^4-r_+^4) \nonumber \\
&&+
\ap {3M^2 \om^2 \V \beta \over 2}\lb{1 \over \tilde R^4
  }-{1 \over r_+^4}\rb.
\een
To calculate $I_1$ we can use the unperturbed solution.
\ben
I_1 &=&{\ap \over 4} \integrat R_{\mu \nu \rho \sigma} R^{\mu \nu \rho \sigma}
\nonumber \\
&=&
\ap \left[ {5 \V \beta \over 2b^4}(\tilde R^4-r_+^4)+{9 M^2 \om^2 V_3 \beta
  \over 2}{1 \over r_+^4}\right].
\een
So, finally we get,
\be \label{eq:r2acnonshell}
I_{BH}=-{\V \beta \over \Nc}\left[-{2 \ob}(\tilde R^4-r_+^4) + \ap{5\over
    3b^4}\Rr +\ap {3M^2 \om^2 \over r_+^4}\right].
\ee
Now let us similarly evaluate the background action for the $AdS$ spacetime,
\be
I_{AdS}=-\nc \lb J_0+J_1 \rb.
\ee
Where,
$$
J_0=-{2 \ob}\beta' \V \tilde R^4 -{5\ap \over 6b^4} \beta' \V \tilde R^4
$$
and,
$$
J_1={5 \ap \over 2b^4} \V \beta' \tilde R^4.
$$
So finally we get,
\be \label{eq:r2bgacnonshell}
I_{AdS}=-{\beta' \V \over \Nc}\left[-{2 \ob}\tilde R^4
+ \ap {5\over 3b^4}\tilde R^4
  \right].
\ee
>From Eqs. \ref{eq:r2acnonshell} and \ref{eq:r2bgacnonshell},
\ben \label{eq:r2subtractacn1}
\Delta I &=& I_{BH} - I_{AdS} \nonumber \\
&&
={\V \over \Nc} \bigg[ {2 \ob} \tilde R^4 (\beta - \beta')-\beta
    {2 \ob}r_+^4 \nonumber \\
&&
-\ap {5 \over 3b^4} \left[\tilde R^4(\beta
      -\beta') -\beta r_+^4 \right]
-\ap \beta{3\om^2 M^2\over r_+^4}\bigg].
\end{eqnarray}
As prescribed, we equate the boundary geometries at
$r={\tilde R}$ to obtain the
relation between $\beta$ and $\beta'$.
\be
\beta' \lb 1+{\tilde R^2 \ob}+\ap {\tilde R^2 \over 6b^4}\rb ^{1/2}=\beta \lb
1+{\tilde R^2 \ob} - {\omega M \over \tilde R^2}
+\ap {\tilde R^2 \over 6b^4} + \ap {M^2
  \om^2 \over 2\tilde R^6}\rb ^{1/2}.
\ee
Therefore,
\be
\label{r2tempreln}
\beta'=\beta \lb 1 -{1 \over 2\tilde R^4} \om M b^2
+ {\ap \over 12 \tilde R^4} \om M \rb.
\ee
Using this relation $\Delta I$ can be written as,
\be
\Delta I=-{\beta \V \over \Nc} \left[ {r_+^4 \ob} -r_+^2 +{\ap \over
    6b^4}\lb 36b^2r_+^2 + 10 r_+^4 +15b^4\rb \right].
\ee
The energy of the black hole is then given by,
\ben
E&=&{\partial \Delta I \over \partial \beta} \nonumber \\
&=&
{\V \over \Nc}\left[{3 \ob}(r_+^4 + r_+^2b^2) -{\ap \over 2b^4} {20
    r_+^6 + 14 r_+^4 b^2 - 6 r_+^2b^4 -3b^6 \over b^2 -2r_+^2}\right]
\nonumber \\
&=& M + O(\ap).
\een
And the Euclidean entropy works out to be,
\be
S={\V r_+^3 \over 4G_5} \left[ 1+ \ap{2 \ob} \lb 1+ {3\over 2}{b^2
    \over r_+^2}\rb \right].
\label{r2seucl}
\ee
Note that though this is to leading order in $\ap$, we have had to
make no assumption on ${r_+^2\over b^2}$.
This answer disagrees with the expression Eq. 70 of \cite{noj}. 
But it agrees with the expressions found in \cite{neupane}.

Using Wald's formula \ref{eq:wald2} the expression for the
entropy is
\be
S_{Wald}={1 \over 4\,G_5}\,\int_{Horizon}d^3x \sqrt{h} \left [ 1+
  {\ap \over 2} \lb R - 2\,h^{ij}\, R_{ij} + h^{ij} \, h^{kl}\,
  R_{ikjl} \rb \right ].
\ee
Evaluating this expression to leading order in 
$\ap$ gives the same answer
Eq.\ref{r2seucl} as the Euclidean calculation.


\subsection{The $R^4$ Term}

Since one of our motivations to study the relation between the
Euclidean and Wald expressions for entropy is the AdS/CFT
correspondence, we should look at the corrections that appear
in the Type IIB string effective action. The first non vanishing
corrections involve eight derivatives -- a term involving four
powers of the Riemann tensor together with its supersymmetric
counterparts.

In the context of corrections to $AdS$ black hole entropy, 
the $R^4$ term was
studied in \cite{gubser}
\footnote{\cite{haro} discusses the effects of these 
higher derivative terms on the extremal D3 brane solution.}. 
They adopted the Euclidean approach and looked at
the so called planar black hole (a large mass limit of the 
$AdS$ Schwarzschild
solution)\footnote{This was generalised to the finite mass case
in \cite{gaoli}\cite{land},
who also used the Euclidean approach.}. They were able to compute
the leading order correction
to the entropy. Here we will study the correction using Wald's
expression and obtain
the same result as \cite{gubser}. We have also further 
checked agreement
with the finite mass results of \cite{gaoli, land}.
\\
$\bt$ {\bf Review of the Euclidean Calculation}

The relevant pieces of the
ten dimensional tree level type IIB superstring effective action are,
\be
I=-{1 \over 16 \pi G_{10}} \int d^{10}x \sqrt{-g_{10}} \lb R - {1 \over
4.5!}(F_5)^2 \rb + I',
\label{tendacn}
\ee
\be
I'= -{\gamma
\over 16\pi G_{10} } \int d^{10}x \sqrt{-g_{10}} W ,
\ee
where, $F_5$ is the self-dual five form field strength. Note that we have set
the dilaton to a constant value $\phi_0$.
Here,
\be
\gamma={1 \over 8} \zeta (3) (\ap)^3 
\ee
and
\be
W=C^{hmnk}C_{pmnq}C_h^{\hhp rsp}C^q_{\hhp rsk}+{1\over
2}C^{hkmn}C_{pqmn}C_h^{\hhp rsp}C^q_{\hhp rsk}.
\ee
where $C_{pqmn}$ is the Weyl tensor. To put the leading
$R^4$ term in this form, one has to use
the freedom of field redefinition of the
metric.

We will consider $I'$ as perturbation.
The leading order Type IIB supergravity solution
which we will study is the throat region of the non-extremal
$D3$-brane.
\be
ds^2={r^2 \over b^2}\left[-\lb 1-{r_0^4 \over r^4}\rb dt^2 +d\vec x^2
  \right]+ {b^2 \over r^2}\lb 1-{r_0^4 \over r^4}\rb ^{-1} dr^2
  + b^2 d\Omega _5^2
\label{nonext10d}
\ee
together with a constant self-dual five form field strength having $N$ units
of flux in the $S^5$
as well as the black hole spacetime.
The radius $b$ is related to the ten dimensional parameters via,
\be \label{eq:adsdictnry}
b^4={N \sqrt{2G_{10}} \over \pi^{2}}.
\ee
Including the eight derivative terms such as $I'$,
the ten dimensional solution Eq.\ref{nonext10d} gets corrected.
However, for the leading order correction, we only need to take into
account the term $I'$.
Moreover, it was argued in \cite{gubser} (see also \cite{theisen}) that
one could compactify the ten dimensional action Eq.\ref{tendacn}
on the $S^5$ to get the effective five dimensional gravitational
action,\footnote{There are some caveats to be mentioned regarding 
the use of this five dimensional effective action. There can be 
higher derivative terms 
involving $F_5$ which might also contribute. See \cite{haro}.
However, our point of view here is more limited and we are simply 
undertaking to check the equality between the computation of 
\cite{gubser} done with this effective action, with Wald's approach.
We thank K. Skenderis for discussion on this point.} 
\be
I_5=-{\nc}\int d^5x \sqrt{-g} \left[\lb R_5+{12 \over b^4}\rb
  + \gamma W \right]
\label{5dacn}
\ee
where,
\be \label{eq:g5g10rel}
 \nc = {Vol(S^5) \over 16 \pi G_{10}}={\pi^3 b^5 \over 16\pi G_{10}}.
\ee
For the computation of the leading correction in $\gamma$
to the entropy, it suffices to
consider this effective action and its solutions. It is consistent
to set the dilaton to its constant value and take the five form to be
self-dual.

The five dimensional leading order solution\footnote{We will use $r_0$
for the location of the horizon of the planar black hole in keeping
with \cite{gubser}, and use $r_+$ for that of the usual finite mass
black hole.}
\be
ds^2={r^2 \over b^2}\left[-\lb 1-{r_0^4 \over r^4}\rb dt^2 +d\vec x^2
  \right]+ {b^2 \over r^2}\lb 1-{r_0^4 \over r^4}\rb ^{-1} dr^2.
\label{5dsoln}
\ee
can be obtained as a large mass limit of the metric Eq.\ref{eq:adssh}.
The temperature is given by
\be
T={r_0 \over \pi b^2}.
\ee
The leading entropy is
\be
S={V r_0^3\over 4b^3G_5}={\pi^2\over 2}N^2VT^3,
\label{leadent}
\ee
where $V=\int d^3\vec x$ is the volume in the dual gauge theory.

The leading correction to the solution, Eq.\ref{5dsoln},
from the Weyl term in Eq.\ref{5dacn}, is given by
\cite{gubser} (in terms of the functions entering the spherically symmetric
ansatz Eq.\ref{eq:sphmet}),
\ben
B(r)={r^2 \over b^2}\lb 1 - {r_0^4 \over r^4}\rb \left [ 1 - {15
    \gamma \over b^6} {r_0^4 \over r^{12}} (5 r^8 + 5 r^4 r_0^4
  -3r_0^8)\right ] \nonumber \\
A(r)= {b^2 \over r^2}\lb 1 - {r_0^4 \over r^4}\rb^{-1} \left [ 1 + {15
    \gamma \over b^6} {r_0^4 \over r^{12}} (5 r^8 + 5 r^4 r_0^4
  -19r_0^8)\right ].
\een
One can then apply
the Euclidean prescription as in the previous subsections.
The final result is
\ben
\label{eq:gktresult}
S &=&S_0\lb 1+{45\gamma \over b^6}\rb
+S_0\lb {15\gamma \over b^6}\rb \nonumber\\
&=& S_0 \lb 1+{60\gamma \over b^6}\rb,
\een
where, $S_0$ is the Bekenstein-Hawking entropy evaluated
on the area of the {\it perturbed} solution
\be
S_0={Area_H \over 4G_5}= {V r_0^3\over 4b^3G_5}.
\ee
The two terms in the first line come from the Einstein-Hilbert term and
the Weyl term in the action respectively.

The correction to the temperature in
the presence of the $R^4$ term is given by,
\be
T={r_0 \over \pi b^2} \lb 1 + {15 \gamma \over b^2}\rb.
\ee
So the
entropy given in Eq.\ref{leadent} is corrected to
\be
S={\pi^2 \over 2} N^2 V T^3 \lb 1 + {15 \gamma \over b^6} \rb.
\label{gktgauge}
\ee

Since Wald's expression gives the area term from the Einstein-Hilbert
Lagrangian, we will get $S_0$
after using the perturbed solution.
Therefore what we will do is to
calculate the correction to the entropy coming from the $R^4$ term
using Wald's formula and show that the sum of the two contributions
matches with the above result.\\
\noindent
$\bt$ {\bf Wald's Formula for the $R^4$ Term}\\
We saw in Sec. 3.3 that the Wald expression for the entropy,
for a Lagrangian which depends polynomially on the Riemann tensor, is given by
\be
S_{BH}= -2\pi \int_{\cal H} d^3 x \sqrt{h}{\partial L \over
\partial \RT} \epsilon_{\mu \nu} \epsilon_{\rho \sigma}.
\ee
$\epsilon_{\mu \nu}$
is the binormal to the bifurcation surface,
normalized such that $\epsilon_{\mu \nu} \epsilon^{\mu \nu}=-2$.
We can take
\be
\epsilon_{\mu \nu}=\xi_{\mu}\eta_{\nu} - \xi_{\nu}\eta_{\mu},
\ee
where $\xi$ and $\eta$ are null vectors normal to the bifurcate killing horizon, with
$\xi \cdot\eta =1$.  We will take $\xi$ to be the killing vector
field,
\be
\xi= {\partial \over \partial t}
\ee
which is null at the bifurcate horizon.
Then $\eta$ can be taken to be
\be
\eta=-f^{-1}{\partial \over \partial t}-{\partial \over \partial r}
\ee
where $f=-{r^2 \over b^2} \lb 1 - {r_0^4 \over r^4} \rb$.
In our particular case, we are interested in the additional contribution
to the entropy from
\be
\Delta L = {\gamma \over 16\pi G_{5}} W.
\ee
Therefore
\be
\Delta S=-{\gamma \over 8G_5}
\int_{\cal H} d^3 x \sqrt{h}{\partial W \over
\partial \RT} \epsilon_{\mu \nu} \epsilon_{\rho \sigma}.
\ee
Since W is function only of the Weyl tensor
\be
C_{abcd}=R_{abcd} + {1 \over 3}
(g_{ad}R_{cb}+g_{bc}R_{ad}-g_{ac}R_{db}-g_{bd}R_{ca})+{1 \over
  12}(g_{ac}g_{bd}-g_{ad}g_{cb})R,
\ee
we may write for $\Delta S$,
\be \label{eq:r4deltas1}
\Delta S=-{\gamma\over 8 G_5} \int_{\cal H} d^3 x \sqrt{h}{\partial W \over
\partial C_{abcd}}{\partial C_{abcd} \over \partial \RT} \epsilon_{\mu \nu}
\epsilon_{\rho \sigma}.
\ee
We then have
\be
{\partial W \over \partial C_{abcd}}= W_1^{abcd} + {1 \over 2}W_2^{abcd}
\ee
where
\ben
W_1^{abcd}&=&g_{li} \bigg [C^d_{\, skj}\,C^{lbcj}\,C^{aksi}
+ C^d_{\, jsk}\,C^{blkc}\,C^{ijsa} \nonumber \\
&&
+C^d_{\, jks}\,C^{ajkl}\,C^{bsic}
+C^d_{\, kjs}\,C^{ljka}\,C^{sbci} \bigg ]
\een
and
\ben
W_2^{abcd}&=& g_{li} \bigg [C^a_{\, vkj}\,C^{ivkb}\,C^{jlcd}
+ C^b_{\, vkj}\,C^{ivka}\,C^{ljcd} \nonumber \\
&&
+C^a_{\, jkv}\,C^{dlkv}\,C^{ibcj}
+ C^d_{\, jkv}\,C^{lavk}\,C^{jbci} \bigg ].
\een
We will denote
\be
M_{abcd}={\partial C_{abcd} \over \partial \RT} (\xi_{\mu}\,\eta_{\nu}-
\xi_{\nu}\,\eta_{\mu})\,(\xi_{\rho}\,\eta_{\sigma} -
\xi_{\sigma}\,\eta_{\rho})
\ee
where
\ben
{\partial C_{abcd} \over \partial \RT} &=&
\delta_a^{\mu}\,\delta_b^{\nu}\,\delta_c^{\rho}\,\delta_d^{\sigma}
\nonumber \\
&&
+{1
  \over 3}\,\lb g_{ad}\,g^{\mu \rho}\, \delta_c^{\nu}\,
\delta_b^{\sigma} + g_{bc}\,g^{\mu \rho}\,
\delta_a^{\nu}\,\delta_d^{\sigma}- g_{ac}\,g^{\mu \rho}\, \delta_d^{\nu}\,
\delta_b^{\sigma}-g_{bd}\,g^{\mu \rho}\, \delta_c^{\nu}\,
\delta_a^{\sigma} \rb \nonumber \\
&&
+{1 \over 24} (g_{ac}\, g_{bd}-g_{ad}\,g_{bc})\,(g^{\mu \rho}\, g^{\nu
\sigma}-g^{\mu \sigma}\,g^{\nu \rho})
\een
Using the unperturbed metric Eq.\ref{5dsoln} one can calculate,
\be
W_1^{abcd}M_{abcd}=-{232 r_0^{12} \over b^6 r^{12}}
\ee
and
\be
W_2^{abcd} M_{abcd}={224 r_0^{12} \over b^6 r^{12}}
\ee
Then the integrand in Eq. \ref{eq:r4deltas1} becomes
\be
{\partial W \over
\partial C_{abcd}}{\partial C_{abcd} \over \partial \RT} \epsilon_{\mu \nu}
\epsilon_{\rho \sigma} \mid_{r=r_0}=-{120 r_0^{12} \over b^6
  r^{12}}|_{r=r_0}=-{120 \over b^6}.
\ee
Together with the factors of $\sqrt {h}={r_0^3 \over b^3}$ and
$V$ in the integral we get
\be
\Delta S={V r_0^3\over 4b^3G_5}\lb {60\gamma \over b^6}\rb
={\pi^2 \over 2}N^2 V T^3 \lb {60 \gamma \over b^6}\rb
\ee
So the total entropy is,
\be
S=S_0 + \Delta S = {V r_0^3\over 4b^3G_5}\lb 1+{60\gamma \over b^6}\rb
={\pi^2 \over 2} N^2 V T^3 \lb1+{15\gamma \over b^6}\rb
\ee
which agrees with Eqs.\ref{eq:gktresult},\ref{gktgauge}.
Note that the individual
contributions from the Einstein-Hilbert term and the
Weyl term to the Euclidean entropy, given
in the first line of Eq.\ref{eq:gktresult}, are
{\it different} from the individual contributions
from these terms to the Wald entropy.

One can also look at the corrections to the \asch \
metric from the action \ref{5dacn} as in \cite{land}.
The Euclidean entropy in this case has been computed in \cite{land}
to be,
\be
\label{largeMentro}
S={\V r_+^3 \over 4 G_5} \bigg [ 1 + {60 \gamma \over b^6} \lb 1 + {b^2
  \over r_+^2}\rb^3 \bigg ].
\ee
In the Wald approach, the Einstein-Hilbert part of the action gives,
$$
S_0={\V r_+^3 \over 4 G_5}.
$$
The correction to the area law comes from the $R^4$ term.
Following the same
procedure as above can calculate the integrand
in Eq. \ref{eq:r4deltas1} using the unperturbed \asch \, metric. The final
result is,
\be
{\partial W \over
\partial C_{abcd}}{\partial C_{abcd} \over \partial \RT} \epsilon_{\mu \nu}
\epsilon_{\rho \sigma} \mid_{r=r_+}=-{120 \gamma \over b^6} \lb 1+ {b^2
\over r_+^2} \rb^3.
\ee
So the total entropy is given by Eq. \ref{largeMentro} and once again the
two approaches yield the same answer for any value of the mass.


\sectiono{\bf Wald's Approach Vs. Euclidean Approach in
Asymptotically AdS Spacetime}

Having seen that the computations in both approaches agree non-trivially in a
number of examples, we have reason to be confident that this must
generally be so.
In this section, we will attempt to gain some understanding of why this
might be so.
We will start with looking at the Noetherian definition
of mass in AdS spacetimes
and see that it involves a subtraction procedure very like
in the Euclidean framework.
We will use this then, together with a line of argument due to
Wald, made for the asymptotically
flat case, to relate the Wald expression for
the entropy to the Euclidean one.

\subsection{Mass in asymptotically AdS spacetime}

There are several ways to define mass in asymptotically AdS
spacetimes (see \cite{marolf} for a comprehensive comparison and
references). In \cite{marolf} it was shown for the Einstein-Hilbert
Lagrangian, with appropriate boundary conditions, that the
Hamiltonian definition of Wald reviewed in Sec. 3.2 (see also
\cite{zoupas}), agreed with several
other definitions. In particular, we have from Eq.\ref{eq:hamq},
\be \delta H_{\xi}=\delta\int_{\tilde R} dS_{ab} \sdg \hps
Q^{ab}[\xi].\label{dqdh}\ee
The boundary term ${\bf\Theta}$ that appears in Eq.\ref{eq:hamq}
does not contribute\footnote{For
simplicity, we will assume in what follows that the boundary term
always cancels out on
subtraction from the background. Which is why we have dropped
it in Eq.\ref{inth}. This is true for the
Einstein-Hilbert Lagrangian and we have checked that it also holds for
the Gauss-Bonnet case. We believe this must be a general feature
in the asymptotically AdS case.
This is {\it unlike} the flat space case where
the boundary term is crucial to the Euclidean computation, being the
only contribution to the mass. In any
case, even if a boundary term contributes, the argument
below can be modified by
including the corresponding term
in the Euclidean computation of the action as well.}.
The Noether charge appearing in the right hand side is well
defined in asymptotically AdS spacetimes only after introducing a cutoff
at an outer boundary (at, say, $r={\tilde R}$).
The variation $\delta$ is then such that it keeps the geometry fixed at
this hypersurface.

In the case, of the Einstein-Hilbert Lagrangian, it was shown in \cite{marolf}
that the boundary conditions allow for
Eq.\ref{dqdh} to be integrated to give a Hamiltonian $H_{\xi}$.
The additive constant to the Hamiltonian is fixed by demanding that
the energy is zero in pure AdS.

We will assume that the above Noetherian
definition of mass continues to make sense
for higher derivative Lagrangians of interest in asymptotically
AdS spacetimes (reflected in appropriate
boundary conditions). Namely, we will integrate Eq.\ref{dqdh}
\be
H_{\xi}=
\left[\int_{\tilde R} dS_{ab} \sdg \hps Q^{ab}[\xi] - \int_{\tilde
R} dS_{ab} \sdg \hps Q_{AdS}^{ab}[\tilde{\xi}]\right].
\label{inth}
\ee
Again, the additive constant has been chosen so that the Hamiltonian is zero
for pure AdS spacetime. Since it shouldn't contribute to the
variation, the boundary geometry at $r=\tilde{R}$ must be the same
for both the spacetime and the reference AdS. This implies that the
killing vector
field $\tilde{\xi}$ is normalised such that it agrees with that of $\xi$
\be
|\tilde{\xi}|^2=|\xi|^2.
\ee
on the boundary hypersurface.

Since $Q[\xi]$ is linear in $\xi$ by construction from
Eq.\ref{eq:noeth2},
and since the difference in normalisation
between $\xi$ and $\tilde{\xi}$ is a constant,
we can write,
\be
Q_{AdS}^{ab}[\tilde{\xi}] = \left [\lb {g_{tt} \over g^{AdS}_{tt}} \rb^{1
\over 2} \right]_{r=\tilde R} Q_{AdS}^{ab}[\xi].
\ee
Here we have taken $\xi$ to be the time translation killing vector.
So,
\be
\int_{\tilde R}
  dS_{ab} \sdg \hps Q_{AdS}^{ab}[\tilde{\xi}] =\left [\lb {g_{tt} \over
      g^{AdS}_{tt}} \rb^{1 \over 2} \right]_{r=\tilde R} \int_{\tilde R}
  dS_{ab} \sdg \hps Q_{AdS}^{ab}[\xi].
\ee

The Hamiltonian which is the total energy or mass of the system
is then given by
\be
{\cal E} =\left[\int_{\tilde R} dS_{ab} \sdg \hps Q^{ab}[t] - \left [\lb
    {g_{tt} \over g^{AdS}_{tt}} \rb^{1 \over 2} \right]_{r=\tilde R}
\int_{\tilde R} dS_{ab} \sdg \hps Q_{AdS}^{ab}[t]\right].
\label{noethmass}
\ee
It is clear from the above expression that for pure AdS space time,
$g_{tt}=g_{tt}^{AdS}$, and hence $H_{\xi}=0$.

In the Appendix we explicitly calculate the energy of the AdS
Schwarzschild metric Eq.\ref{eq:adssh} using this
prescription. We see there that the subtraction plays exactly
the same role as it did in the Euclidean computation. It is
necessary for correctly getting the finite answer.

\subsection{Relating the Wald and Euclidean approaches}

Using the above definition of mass, we can make a direct connection
between the Wald and
Euclidean approaches by modifying an argument given by
Wald \cite{wald} for the asymptotically flat case.

As in Sec.3.3 we will consider $\xi$ to be a killing vector vanishing
on the bifurcate horizon. In that case ${\cal L}_{\xi} \psi = 0$ 
and hence ${\bf
\Theta({\cal L}_{\xi} \psi)}=0$. Therefore the Noether current simplifies
to
\be\label{waldeuc}
{\bf J} = -\xi\cdot {\bf L}.
\ee
Integrating both sides of Eq.\ref{waldeuc}
over a constant time hypersurface ${\cal C}$ of the black
hole spacetime, having the
interior boundary {$\cal H$} and the outer boundary at $r={\tilde R}$. We get
$$
\int_{{\cal C}} dV_t J^t = -\int_{\cal C} dV_t \xi^t L(g_{BH})
$$
$$
-\int_{{\cal H}}dS_{ab} \sdg \hps Q^{ab} +\int_{\tilde R}dS_{ab} \sdg \hps
  Q^{ab} = -\int_{\cal C} dV_t \xi^t L(g_{BH}).
$$
Therefore
\be
\int_{{\cal H}}dS_{ab} \sdg \hps Q^{ab}[\xi^t]={\cal E}
+\int_{\cal C} dV_t \xi^t L(g_{BH})+\left [\lb
    {g^{BH}_{tt} \over g^{AdS}_{tt}}
\rb^{1 \over 2} \right]_{r=\tilde R} \int_{\tilde R}
  dS_{ab} \sdg \hps Q_{AdS}^{ab}[\xi^t].
\ee
Using the same logic as above, but now integrating over
$\Sigma$ a Cauchy hypersurface in global AdS, we have
\be
\int_{\tilde R} dS_{ab}
\sdg \hps {Q}_{AdS}^{ab}[\xi^t]= -\int_{\Sigma} dV_t \xi^tL(g_{AdS}),
\ee
As a result,
\be\label{eucnoeth}
\int_{{\cal H}}dS_{ab} \sdg \hps Q^{ab}={\cal E}
+\int_{\cal C} dV_t \xi^tL(g_{BH})
-\lb{g_{tt} \over g^{AdS}_{tt}}\rb^{1 \over 2}\int_{\Sigma} dV_t
\xi^t L_{AdS}.
\ee
Since the Wald entropy of the black hole is
\be
S=\beta\int_{{\cal
    H}}dS_{ab} \sdg \hps Q^{ab},
\ee
Eq.\ref{eucnoeth} becomes
\be
S=\beta {\cal E} +\beta \int_{\cal C} dV_t \xi^tL(g_{BH})-
\beta \lb{g_{tt} \over g^{AdS}_{tt}}\rb^{1 \over 2}\int_{\cal C} dV_t \xi^t 
L(g_{AdS}).
\ee
Since, we have a static background
\be
I_{BH} = -\beta \int_{\Sigma} dV_t \xi^tL(g_{BH})
\ee
and
\be
I_{AdS}= -\beta \lb{g_{tt} \over g^{AdS}_{tt}}\rb^{1 \over 2} \int_{\cal C} 
dV_t \xi^t L(g_{AdS})
\ee
provided we assign a temperature $\beta'$ to $AdS$ space which is
\be
\beta'=\beta \lb{g_{tt} \over g^{AdS}_{tt}}\rb^{1 \over 2}.
\ee

We now see exactly the Euclidean prescription, where one subtracts the action
of a background  AdS spacetime with the above identification of temperatures.
In other words,
\be
\label{thermrel}
S=\beta {\cal E} - I_{BH} + I_{AdS} =\beta {\cal E} - \Delta I.
\ee
Thus starting from the Noetherian expressions for the entropy, we obtain
the relation to the Euclidean prescription with exactly the same
subtraction procedure.\footnote{\cite{skenderis} derives
this relation for the general two derivative lagrangian in asymptotically
locally $AdS$ spaces. See also footnote 3.}

Though the mass that appears above is the Noetherian definition of mass,
it must be that it agrees with the Euclidean prescription for the mass.
This is because both Euclidean and Noetherian prescriptions obey the
first law. In appendix B we indicate the argument.

\sectiono{Conclusions}

We have studied the relation between the Euclidean and Noetherian
approaches to the entropy of asymptotically $AdS$ black holes with the
aim of shedding light on the two holographic descriptions for black hole
entropy. The agreement between these two approaches can be understood,
as we have described, from the general construction of Noether charges.
The explicit computations in a number of examples
further bolsters the case for the equivalence of the two approaches.

What would be nice to see is if the argument for this equivalence
can be cast in a way which makes the relation of the Wald's formula
to the gauge theory more transparent. It would help us answer
the question: {\it What is the meaning of Wald's formula
in the dual Gauge Theory?} Given the generality of Wald's formula,
it seems likely that there is an equally universal statement to be made in
the dual gauge theory. It has
presumably to do with the behaviour of the number of
degrees of freedom under RG flow in the gauge theory. The boundary
holographic description (which is related to the Euclidean approach)
is naturally an expansion about small gauge coupling, moving inwards from
the UV so to say.
While the Wald expression
is an expansion in inverse powers of the gauge coupling,
systematically moving outwards from the IR. In this context,
perhaps a generalisation of the
entropy function of Sen \cite{sen1,sen2}
might be a useful way to
understand the interpolation.

Another point which is worth noting from the explicit results we have
exhibited is the relative computational simplicity of the Wald approach
in evaluating corrections. It seems to be less onerous than the Euclidean
procedure. This is also a sign that the Wald approach is more natural, 
at least for large gauge coupling.
\\
\\
\noindent
{\large \bf Acknowledgment}\\
\\
We would like to thank D. Astefanesei and especially A. Sen
for helpful comments during the course of this work. We also
wish to thank K. Skenderis for very useful correspondence.
We are also indebted to the people of India for their support.
\newpage
\noindent
{\Large \bf Appendix}
\appendix
\sectiono{Calculation of Energy}

Let us calculate the total energy for the AdS-Schwarzschild space
time
\be
ds^2=-\lb 1+{r^2 \over b^2} - {\omega M \over r^2}\rb
dt^2 + \lb 1+{r^2 \over b^2} - {\omega M \over r^2}\rb^{-1}dr^2 +
r^2 d\Omega_3^2
\ee
with the Einstein-Hilbert Lagrangian
\be
L= \nc \lb R + {12 \over b^2} \rb.
\ee
For this Lagrangian,
\be
Q^{ab}=-\nc \lb
\nabla^a\,\xi^b\,-\,\nabla^b\,\xi^a \rb.
\ee
Let $\xi$ be an asymptotic
time translational vector,
\be
\xi = {\p \over \p t},\, \,
\,\xi^t=1.
\ee
So
\be
Q^{tr}=\nc \, \,  (\p_r
g_{tt}) \, \xi^t.
\ee
For the AdS-Schwarzschild metric,
\be
Q^{tr}={2 \over \Nc} \lb {r \over b^2}\, + \, {\omega M \over r^3}
\rb.
\ee

The Noetherian definition of mass Eq.\ref{noethmass} is
\be
{\cal E}=\left[\int_{\tilde R} dS_{ab}
\sdg \hps Q^{ab}[t]-\int_{\tilde R}
  dS_{ab} \sdg \hps Q_{AdS}^{ab}[\tilde{t}]\right].
\ee
Now,
\be
\int_{\tilde R} dS_{ab} \sdg \hps
Q^{ab}[t]=\int_{\tilde R} d{\bf S_{tr}} \hps Q^{tr}[t]
\ee
where
${\bf dS^{ab}}= {1 \over 2}(dx^{a} \bigotimes dx^{b} - dx^{b}
\bigotimes dx^{a}) \sqrt{h}$
is the volume element on the boundary.
So putting every thing together we get,
\be
\int_{\tilde R} dS_{ab}
\sdg \hps Q^{ab}[t]={2V \over \Nc} \lb {\tilde R^4 \over b^2} +
\omega M \rb = {2V \over \Nc} {\tilde R^4 \over b^2} +{2M \over 3}.
\ee
Let us calculate the contribution from the background AdS metric,
\be
ds_{AdS}^2=-\lb 1+{r^2 \over b^2} \rb dt^2 + \lb 1+{r^2 \over
b^2}\rb^{-1}dr^2 + r^2 d\Omega_3^2.
\ee
Here
\be Q_{AdS}^{tr}={2
\over \Nc} {r \over b^2} \, \tilde{\xi}^t
\ee
where the asymptotic time
translation vector $\tilde{\xi}= \tilde{\xi}^t {\p \over \p t}$ in
background AdS space is given by,
\be
\tilde{\xi}^t=\lb{g_{tt} \over
g_{tt}^{AdS}}\rb^{1 \over 2}=1-{\omega
  M  b^2 \over 2 \, \tilde R^4}.
\ee
Therefore
\be
\int_{\tilde R} dS_{ab} \sdg \hps
Q_{AdS}^{ab}[\tilde{\xi}^t]={2V \over \Nc}
    {\tilde R^4 \over b^2}\lb 1-{\omega
  M  b^2 \over 2 \, \tilde R^4} \rb
={2V \over \Nc} {\tilde R^4 \over b^2}-{M \over 3}.
\ee
This finally implies
\be
{\cal E} = M.
\ee


\sectiono{ Relating the Noetherian and Euclidean definitions of Mass}

In this section we will show that the Euclidean definition of mass is
the same as that given by the Noetherian method. The
Euclidean definition of mass is given by,
\be
M={\p \Delta I \over \p
\beta}
\ee
where $\Delta I=I_{BH}-I_{AdS}$,
\be
I_{BH}=-\beta \, \int_{\cal C} \, \xi \cdot {\bf L}, \, \, \, \, \,
\,\, \, I_{AdS}=-\beta_{AdS} \, \int_{\Sigma} \, \xi \cdot {\bf L_{AdS}}.
\ee
Using Eq.\ref{waldeuc}, $I_{BH}$ and $I_{AdS}$ can be written as,
$$
I_{BH}=\beta \int_{\tilde R} Q[\xi]- \beta \int_{{\cal H}} Q[\xi]  
$$ and
$$
I_{AdS}=\beta \lb {g_{tt} \over g_{tt}^{AdS}}\rb^{1/2}
\int_{\tilde R} Q_{AdS}[\xi]
$$
So,
\ben
{\p \Delta I \over \p \beta} &=& \int_{\tilde R} Q[\xi] -\lb {g_{tt} \over
  g_{tt}^{AdS}}\rb^{1/2} \int_{\tilde R} Q_{AdS}[\xi]
-{\p \over \p \beta} \lb \beta \int_{{\cal H}} Q[\xi] \rb +\beta {\p
{\cal E}\over \p \beta} \nonumber \\
&&
= {\cal E} -\Bigg [{\p \over \p \beta} \lb \beta \int_{{\cal H}} Q[\xi] \rb
-\beta {\p {\cal E} \over \p \beta} \Bigg ]
\een
In the first line (as well as to go to the
second line) we have used the Noetherian definition of mass
Eq.\ref{noethmass}.
Since the Wald's expression for the entropy obeys the first law, the term
inside the bracket vanishes.
We therefore get
\be
{\p \Delta I \over \p \beta}={\cal E}.
\ee




\end{document}